\documentclass[12pt]{article}
\usepackage{epsfig}
\usepackage{amsmath}

\def\apj {ApJ}
\def\prd {Phys. Rev. D}
\def\prc {Phys. Rev. C}
\def\mnras {Mon. Not. Roy. Astro. Soc.}

\def\beq{\begin{equation}}
\def\eeq#1{\label{#1}\end{equation}}
\def\eeqn{\end{equation}}

\def\beqa{\begin{eqnarray}}
\def\eeqa#1{\label{#1}\end{eqnarray}}
\def\eeqan{\end{eqnarray}}

\let\bar=\overbar

\def\O{{\cal O}}

\def\Dslash{\not{\hbox{\kern-4pt $D$}}}
\def\dslash{\not{\hbox{\kern-2pt $\del$}}}

\def\msb{{\bar{\ssstyle M \kern -1pt S}}}

\usepackage{fancyhdr,graphicx}
\def\Title#1{\begin{center} {\Large {\bf #1} } \end{center}}

\def\codename{Burn-UD}

\begin{document}


\Title{Explosive Combustion of a Neutron Star into a Quark Star: the non-premixed scenario}

\bigskip\bigskip


\begin{raggedright}

{\it Rachid Ouyed $^1$, Brian Niebergal $^1$ and Prashanth Jaikumar $^2$\\
$^1$Department of Physics and Astronomy, University of Calgary, Calgary, Alberta,  T2N 1N4,
Canada\\
$^2$Department of Physics and Astronomy, California State University,  Long Beach
1250 Bellflower Blvd., Long Beach, CA 90840-9505, USA\\~\\
 {\tt Email: rouyed@ucalgary.ca}}
\bigskip\bigskip
\end{raggedright}

\section{Introduction}
\label{sec:introduction}

The traditional strange-quark matter (SQM) hypothesis is the assertion that
bulk matter (large baryon number) comprised of deconfined \textit{up, down, \& strange} quarks (or more conveniently written throughout the rest of this paper as \textit{u}, \textit{d}, \& \textit{s}) at zero pressure is more energetically favorable than the most stable atomic nuclei \cite{1971PhRvD...4.1601B,1979PThPh..61.1515T,PhysRevD.30.272}. Thus, once formed, SQM may exist forever.  Nuclei are then only a metastable state. This hypothesis can be described  using the MIT bag model - a simple model that nevertheless has been reasonably successful in describing thermodynamical properties of quark matter (see \cite{1996csnp.book.....G} and references therein).

In this paper, we review our numerical investigation of  the issue of combustion of pure neutron matter to \textit{u,d,s} matter.
For SQM burning, far from turbulence-flame interactions, even the laminar flow has not been analyzed in sufficient detail. An improved prescription of the burning front in the laminar flow approximation was given in \cite{2010PhRvC..82f2801N,Niebergal-thesis-2011}. Our simulations use hydrodynamics, taking into account binding energy release and neutrino emission across the burning front  - going beyond previous treatments of the problem. Our studies already find speeds as high as $\sim c/100$, where $c$ denotes speed of light. We find  indications that a de-leptonizaton instability (a new type of instability unique to this type of 
conversion) and probably turbulent effects  may well decide the fate of the conversion (deflagration or detonation). 

 This problem is interesting for three main reasons: (i)
in Type Ia supernovae, multi-dimensional studies of small scale dynamics of flame burning including turbulence provide support for a pathway to the distributed regime, which can be a platform for detonation of the white dwarf~\cite{2008ApJ...689.1173A,2009ApJ...704..255W,2008ApJ...681..470P}. A delayed detonation model for Type Ia explosions has also been discussed in the context of material unburnt by the initial deflagration~\cite{2005ApJ...623..337G}. Furthermore, deflagration-to-detonation (DDT) transition through turbulent flame burning in laboratory experiments on combustible gas mixtures has been numerically studied~\cite{Khokhlov:1999} - all this naturally leads to an investigation of the existence of similar effects in SQM burning (ii)
recent work on burning neutron matter to SQM~\cite{2010arXiv1005.4302H} uses simple scaling arguments to show that the above pathway to explosive detonation may occur for SQM burning in a neutron star and (iii) explosive conversion of a neutron star to a strange quark star is astrophysically relevant as discussed in this paper.

Following this very  brief introduction, the rest of the paper is structured as follows: In Section 2,  we present results from numerical simulations  of the combustion from hadronic to quark matter and provide a recipe on how
to set up such simulations.  This is essentially an investigation on how hadronic matter would transition to \textit{u,d,s}-quark matter
through the conversion of hadrons to \textit{u,d}-quarks, then \textit{u,d}-quarks to \textit{s}-quarks.
We also present an analysis of the similarities between
this transition and a combustion process, specifically a combustion of the \textit{non-premixed} type, a distinction that is critically important\footnote{Previous literature on this topic has reported very different results on the transition speed and energy as a result of (incorrectly) assuming \textit{premixed} combustion}. A discussion of the results and future outlook for numerical studies of the SQM burning are presented in Section 3 before we conclude in Section 4.

\section{Simulations of The Hadronic-to-Quark-Matter Phase Transition}
\label{sec:simulations}

Although asymptotic freedom would appear to suggest otherwise, even at supra-nuclear densities (i.e, higher than that of a nucleus),
nucleons cannot spontaneously dissolve into their constituent 
{\it u} \& {\it d}-quarks as the overall energy per baryon would be higher.
The presence of \textit{s}-quarks adds an extra (flavor) degree of freedom, 
thus lowering the overall energy per baryon \cite{1971PhRvD...4.1601B};
resulting in the SQM phase of matter. Since the probability of converting many \textit{u}- \& \textit{d}-quarks into \textit{s}-quarks at the same time
is effectively zero, a conversion from hadronic to SQM can only proceed in practice if a substantial number of \textit{s}-quarks are present.

Consider the region inside a neutron star (NS) where the central density has reached that of nuclear deconfinement. Subsequently, a seed of SQM is instantiated at the star's center. It has been speculated that such a situation could occur during the core-collapse phase of a supernova, or, if the nucleation timescale is slower~\cite{1994PhRvD..50.6100L}, in an older NS whose central density has increased due to spin-down \cite{2006ApJ...645L.145S}.
The details of the seeding mechanism are not considered here but there are several possibilities, for example clustering of lambdas, higher-order neutrino ``sparking'' reactions, or seeding from the outside \cite{1986ApJ...310..261A}.

\begin{figure}[t!]
\centering
\includegraphics[width=0.8\textwidth]{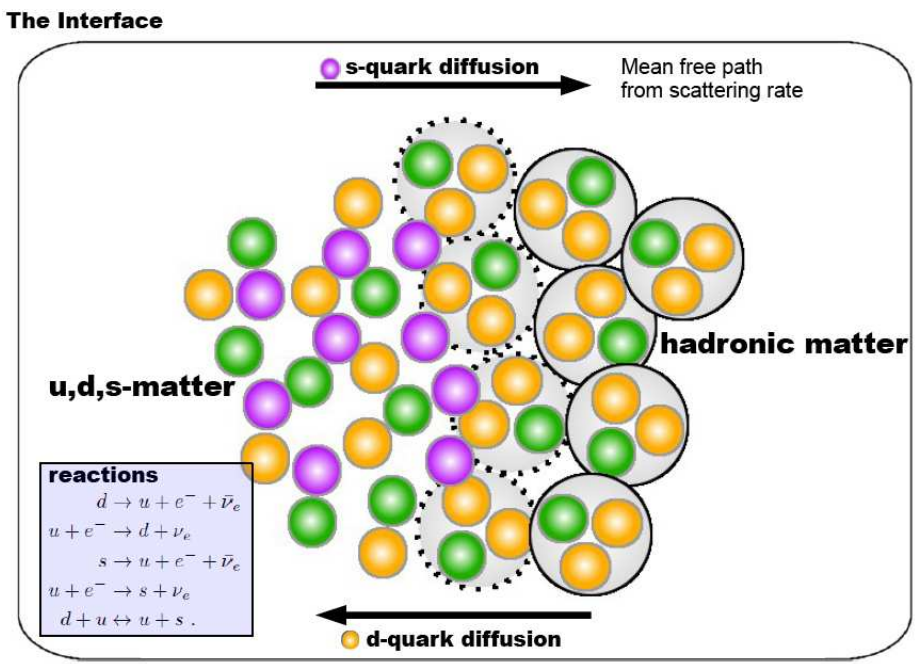}
\caption[Interface]
{\label{fig:interface} The interface where hadronic matter is one side and SQM on the other, with the interplay between  reactions and diffusion governing the speed at the which the interface spreads. The 
presence of $s$-quarks activates the burning and so the process is governed by particle diffusion rather than heat diffusion.}
\end{figure}

In the boundary region surrounding the SQM seed, hadrons
will then overlap with \textit{s}-quarks, allowing the hadrons 
to dissolve into their constituent \textit{u}- \& \textit{d}-quarks. 
Such a region will attempt to equilibrate chemically by producing more \textit{s}-quarks, which will diffuse into the hadronic regions, forming even more SQM. The situation is then an interface where hadronic matter is on one side and SQM on the other, with the interplay between reactions and diffusion governing the speed at which the interface spreads (see Figure \ref{fig:interface}). This scenario is of a type similar to that of \textit{non-premixed combustion} \cite{2000tuco.book.....P},
rather than \textit{premixed}. The important difference is that the reactions in question are not activated above a certain temperature, as is often the case for combustion processes.  Instead, the presence of \textit{s}-quarks activates the burning and so the process is governed by particle diffusion rather than heat diffusion. If one uses only the jump conditions, one is assuming that only fluid pressure differences matter - this is like setting fire to a barrel of oil, where everything is flammable on contact. However, we take into account diffusion of strange quarks, starting from a non-premixed state, and this slows down the burning speed, but not by more than an order of magnitude surprisingly. Further,  neutrinos  change the temperature profile at the interface and alter the jump conditions quite a bit.
Further issues regarding the nuances of the burning regime are discussed in \S~\ref{sec:combustion_turbulence}.

During the combustion an interface is created with cold (unburnt) fuel on one side and hot (burnt) ash on the other.  Subsequent cooling of the ash will result in pressure gradients that induce fluid-like motion of the matter in and around the interface. To conserve energy, fluid velocities across the interface would then differ, causing compression or rarefaction in the burning region which can enhance or quench the combustion. Thus, a numerical fluid-dynamical treatment of the problem is appropriate, even essential, to determine the interface speed.

This problem was addressed numerically in one dimension by \cite{2010PhRvC..82f2801N}.
 We refer the interested reader to 
 \cite{Niebergal-thesis-2011} where 
 the problem is first formulated in a reactive-diffusive hydrodynamic setup (see \S~4
 in  \cite{Niebergal-thesis-2011}). The importance of conserving energy and generating entropy while re-arranging particles into different states is emphasized from the start. Details of the numerical procedure and relevant techniques are given in \S~5 in  \cite{Niebergal-thesis-2011} and instructions are included for obtaining and running \codename.  
In \S~6 in \cite{Niebergal-thesis-2011}, basic results of \codename~  such as the combustion front speed are verified with analytic solutions (jump conditions).
The broader implications of the numerical and analytic results are described in \S~7 in \cite{Niebergal-thesis-2011}, with applications to compact stars, relativistic heavy-ion collision experiments, and the early Universe.   
Finally, animations of the velocity and temperature profiles, as well as the reaction rate, neutrino emissivity and mean free paths
can be found online at:   ``http://quarknova.ucalgary.ca/software/Burn-UD/"

The most notable results of the modern numerical approach are the rapid burning speeds (0.002-0.04 times the speed of light) and the quenching of the burning due to neutrino cooling, resulting in a new type of instability
 as discussed in  \S \ref{sec:new_instability}.

\subsection{Evolution of the chemical potentials}

\begin{figure}[t!]
\centering
\includegraphics[width=0.8\textwidth]{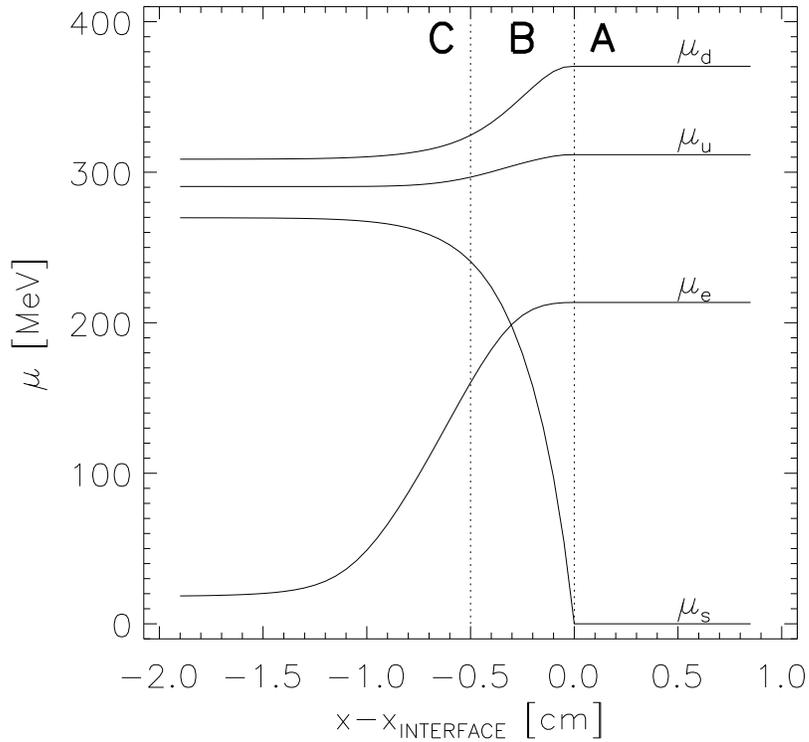}
\caption[A snapshot of the chemical potentials throughout the burning interface.]
{\label{fig:rxn-dfsn_profiles} A snapshot of the chemical potentials for different species 
(\textit{u,d,s} quarks, and electrons) throughout the burning interface.  
Meta-stable \textit{u,d} matter (fuel) is upstream of the interface (zone A). 
Upon contamination by diffusing \textit{s} quarks (oxidizer), the non-leptonic reaction (eq. 4.15 in \cite{Niebergal-thesis-2011};
eq. 6 in \cite{2010PhRvC..82f2801N})
burns the \textit{d} quarks into \textit{s} quarks (zone B).  
In zone C the leptonic reactions (eqns. 4.11-4.14 in \cite{Niebergal-thesis-2011}; eqns 2-5 in \cite{2010PhRvC..82f2801N}) become the dominant
means of producing \textit{s} quarks, leading finally to equilibrated SQM.
The relevant initial parameters are density $\mu_{\rm INIT} = 300$ MeV and electron fraction $\chi_e = 0.04n_{\rm TOT}$.
Leaving hydrodynamics out of the simulations has little effect on these abundance profiles  but enlarges the width of the interface by a factor of $\sim 10$.}
\end{figure}

The spatial profile of the different particle species is shown in Figure \ref{fig:rxn-dfsn_profiles}.
Just after contamination by \textit{s}-quarks, in zone B (cf. Fig. \ref{fig:rxn-dfsn_profiles}), at small values of $\mu_s$, 
the reactions producing \textit{s}-quarks are dominated by the $(\mu_d-\mu_s)^3$ factor in equation 7 in \cite{2010PhRvC..82f2801N} (or
eq. 4.16 in \cite{Niebergal-thesis-2011}).
Further behind the interface, \textit{s}-quark production becomes increasingly 
dependent on the temperature term rather than the chemical potential difference, which tends to increase the burning speed.  
This importance of temperature on the burning process has also been realized in analytical studies \cite{1987PhLB..192...71O,1991NuPhS..24..144H}.

The \textit{s}-quark mass value does have a mild effect on the burning speed in the following way. The front speed, as we will show in \S\ref{sec:new_instability}, is tempered by neutrino cooling, which in our simple treatment of neutrino transport, depends mainly on the neutrino mean free path due to scattering on electrons. The electron density drops as the fuel (\textit{u,d}-quark matter) is burnt.
A larger value for the \textit{s}-quark mass leaves behind a greater electron fraction, resulting in less neutrinos leaving
the system.  This implies that the interface will move faster.
however, this effect is negligible compared to the theoretical uncertainties in the neutrino cooling rate. Thus, changing the \textit{s}-quark mass, or the initial electron fraction, modifies the width and speed of the interface only by a few percent. 

The initial temperature also plays a minor role, provided the assumption of cold degenerate matter ($T\ll\mu\sim 300~{\rm MeV}$) is maintained. Hence, an independence on all initial parameters, except density, is exhibited by the system.

The bag constant was kept fixed even though the behavior of the interface depends upon its value. The justification is that the bag constant may have ill-described dependencies on other variables, for example the chemical potential or temperature \cite{2002PhLB..526...19B}. 
A more self-consistent study then would be one that, rather than varying the bag constant, explores different equation-of-states.

\subsection{A New Type of Instability: wrinkling due to de-leptonization}\label{sec:new_instability}

\begin{figure}[t!]
\centering
\includegraphics[width=0.8\textwidth]{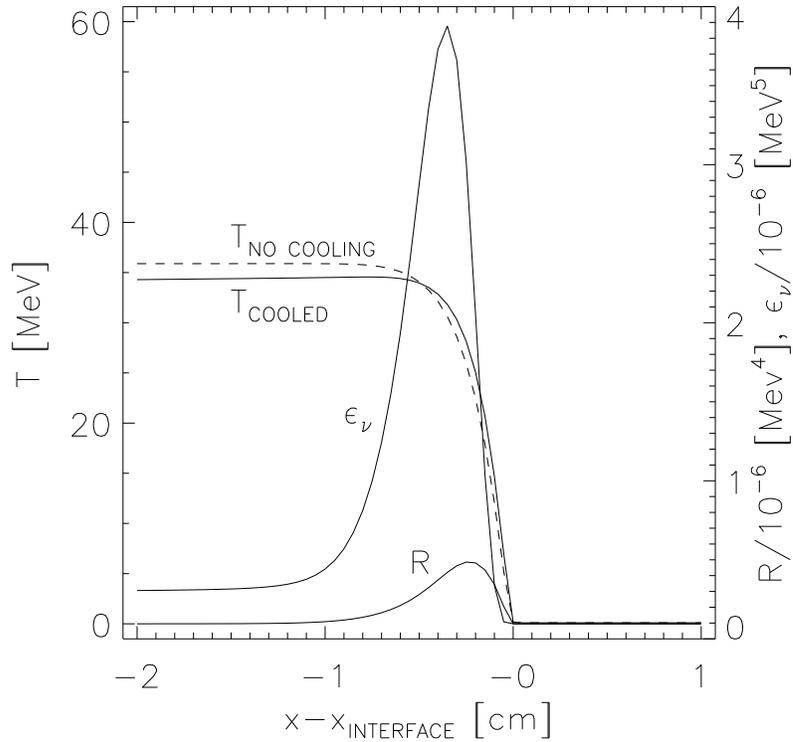}
\caption[A snapshot of the chemical potentials throughout the burning interface.]
{\label{fig:rxn-dfsn_profiles2} A snapshot during the simulation of the temperature
($T$ ), reaction rate ($R$), and neutrino emissivity, $\epsilon_{\nu}$, throughout
the burning interface. The temperature is shown with (solid
line) and without (dashed line) neutrino cooling effects, where
the difference between the two is the variable $C = T -T_{\rm cooled}$
that serves as the measure of cooling.}
\end{figure}

The inclusion of neutrino cooling in the simulations, through the production and transport of neutrinos (as described in \S 4.1.3 in \cite{Niebergal-thesis-2011}; see also \cite{2010PhRvC..82f2801N} ), results in a reduced pressure for the burnt fluid.  This gradient forces a decrease in fluid velocities both behind ($v_2$) and in front ($v_1$) of the interface, causing advection to oppose the progression of the burning interface.
Hence, as the neutrino cooling rate increases the burning speed decreases.
An analytic treatment using the jump conditions (energy-momentum conservation) confirms this effect and also predicts that, for a given density there is a critical neutrino-cooling rate that quenches the burning \cite{2010PhRvC..82f2801N}.

Above the critical cooling rate, instead of moving backwards the interface comes to a halt, since as the interface stops diffusing into the fuel, reactions no longer proceed and neutrino production ceases. Thus, energy is no longer being removed from the burnt region, which causes the pressure gradients to weaken.
The system then reaches a situation where diffusion and advection are in balance, and the interface is stopped.

In other words, the velocity of the (backwards-moving) fluid ahead of the interface is proportional to the neutrino cooling, unless it
becomes larger than the reactive-diffusive burning speed\footnote{The reactive-diffusive burning speed, $v_{RD}$, is the burning speed in the absence of any fluid velocities.} $v_{\rm RD} < \left|v_1\right|$;
at this point it loses its dependence on neutrino cooling and becomes constant.
Only at lower densities ($\leq 2$ times nuclear saturation density) is the interface moving slow enough to allow the burnt material to cool sufficiently and halt. Otherwise the neutrino cooling rate would need to be unphysically large to halt the combustion.

Consider now the interface in more than one dimension.
As a few regions along the burning interface are halted due to cooling, unburnt material starts to flow backwards onto the interface, whereas regions not halted by cooling will be pushing unburnt material away from the interface. The burning interface is mostly opaque to neutrino cooling, but not entirely. ÊThere is a small drop in temperature behind the interface (see Figure \ref{fig:interface}). ÊSince Pressure goes as $T^4$ this is significant. ÊThe burnt areas end up underpressured, which halts the progression of the interface in that region. ÊThe neighbouring regions are still advancing, which creates severe wrinkling, which increases the burning speed. The result is a wrinkled interface, with a shearing between the unburnt fluids (fuel) in halted and non-halted regions.
Since a wrinkled interface increases the diffusion rate,  the effect is an overall increase in the burning speed.  Thus, the non-halted regions
will tend to burn even faster, thereby increasing the wrinkling - leading to an instability.

The wrinkling will be subject to stabilization - in the dimension along the interface diffusion causes concave regions to accelerate while convex regions are decelerated.  Whether this stabilization is able to overcome the instability, or, if the instability grows unhindered is an open question requiring high-resolution multi-dimensional simulations. In any case there is a new type of instability which, due to deleptonization, can halt the
combustion of \textit{u,d}-quark matter.

\subsection{Combustion of a Neutron Star}
\label{subsec:combustion}

In the case of a NS burning into \textit{u,d,s}-quark matter, there is a radial density gradient in the star that the burning interface progresses through. As the interface burns outwards it reaches lower densities, making the interface move slower, until it reaches the critical point where the interface halts due to the above-mentioned deleptonization process.  
Note that, the critical point depends on a number of factors including the equation-of-state, neutrino cooling efficiency, and most importantly the dynamics of the combustion in multiple dimensions  (eg. the wrinkling of the combustion front, sensitivity to instabilities, etc.).

\subsubsection{Low-Pressure Cavitation}

\begin{figure}[t!]
\centering
\includegraphics[width=0.8\textwidth]{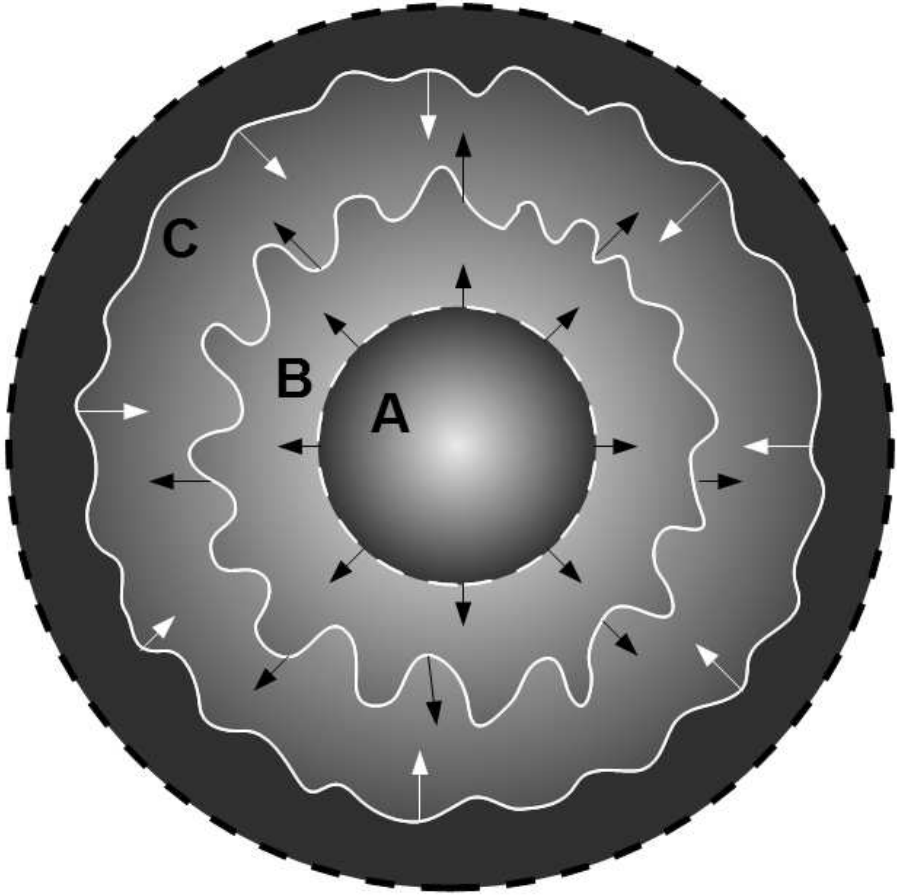}
\caption[Depiction of the burning of a NS.]
{\label{fig:diagram} An illustration of the burning of a NS.  Region \texttt{A} denotes the initial burning, where the interface is laminar and moving outwards at $10^{-3}$ to $10^{-2}$ times the speed of light.
In region \texttt{B} instabilities - such as the Rayleigh-Taylor instability - are likely to develop, causing the interface to wrinkle
and proceed even faster.  At this point either the interface will detonate and eject the outer layers of the original NS, or,  if the wrinkling is not strong enough, the interface will be subject to the deleptonization instability (see \S\ref{sec:new_instability}) and halt (region \texttt{C}).  
The outer layers of the original NS will then fall onto the newly
created \textit{u,d,s}-quark core, leading to a core-collapse and subsequent core-bounce. In either case a highly-explosive event is realized.}
\end{figure}

Eventually the entire interface will reach low density regions and halt. ÊThe core will be underpressured, and the outer material will be falling into it (being burnt as it does). The situation is then an under-pressured \textit{u,d,s}-quark core, with the outer layers of the NS lying on top. The under-pressured core will eventually collapse - similar to the core collapse phase during a supernova - causing the outer lying material to fall onto the core, leading to a second explosion\footnote{The first explosion being the supernova that created the NS.}.  
This sequence of events is illustrated in Figure \ref{fig:diagram}.

\subsubsection{Combustion Regime, Turbulence, \& Jump Conditions}
\label{sec:combustion_turbulence}

As noted above, the burning regime is that of \textit{non-premixed combustion} \cite{2000tuco.book.....P}, rather than \textit{premixed}.  Although this distinction may seem trivial it has important consequences,
such as invalidating assumptions made in the literature for turbulent burning, specifically the flamelet-regime approximation.
We expect turbulence to have a significant impact on the wrinkling of the combustion interface. This will increase the burning speed, not only due to an increase in the surface area of the combustible materials, but also due
to a significant increase in the diffusion. An important point to realize is the effect of temperature/cooling on the combustion to quark matter. Not only can cooling change the properties of the interface such as width and speed, but can quench the burning process entirely. Thus, it is essential to include temperature increase when solving the jump conditions, as was done in \cite{2010PhRvC..82f2801N}. There are numerous examples in the literature where the jump conditions are solved and the temperature of the SQM phase was set to zero! On the issue of turbulence, we remark that Herzog and R\"opke \cite{HR2011}  recently presented results on the burning of neutron matter to SQM including turbulent effects such as Rayleigh-Taylor instabilities. This is distinct from the neutrino-driven instability we mention above. Qualitatively, they too find that
the burning front stops because the reactions are no longer exothermic. However, this ignores the possibility of having a transition to a detonative regime driven by the neutrino-driven instability.

\subsubsection{Transition to Superconducting Phases}

In a manner similar to Cooper pairing of electrons, weakly interacting quarks at high density are expected to pair in order to minimize their free energy, 
resulting in a color-superconducting state (e.g. the Color-Flavor Locked  phase; \cite{PRL86}). This superconducting phase transition is expected to occur in quark stars shortly after they are formed,
since cooling processes are very efficient in lowering the star's temperature
below the rather large critical temperature for color superconductivity $T_c \sim 10$ MeV. The effects of color superconductivity can be taken into account upon the formation of a superconducting interface, where on the superconducting side an extra term ($\propto \Delta$ the pairing energy gap) would have to be included in the equation-of-state.  The \codename ~could then be used to study the dynamics of this interface. Care would need to be taken in order to ensure electrodynamic constraints across the interface are satisfied 
(further details are given in Part III  in  \cite{Niebergal-thesis-2011}).

Furthermore, in one of these phases, the Color-Flavor Locked (CFL) phase,  the photon emissivity dwarfs the neutrino emissivity \cite{Vogt2004,2005ApJ...632.1001O}. 
In this situation matter at a certain radius in the star\footnote{It is not clear whether the inner or outer layers of the star
will drop below the critical temperature first} will drop below $T_c$ causing matter in this region to release energy - (photons and neutrinos) freed as a result of the decay and interaction of low-energy (scalar and vector) modes.  
One can obtain a very high number density of thermal photons (photon fireball)
that can drive the outer layers of the NS like a piston, with important consequences for heavy-element nucleosynthesis \cite{Jaikumar:2006qx}.

\section{Discussion and Outlook}
\label{sec:discussion}

\begin{itemize}

\item {\bf Nucleation}:  A more detailed picture of the pathway to the quark-hadron phase transition is required if we are to establish astrophysical implications of the transition. Basically, one needs to know how long it takes for a seed of quark matter to nucleate and grow inside the NS before it can be analyzed as a ÔbulkÕ fluid (as was implicitly assumed in the combustion studies). Knowing the nucleation time accurately is critical. For example, in several long GRBs, an association with a prior supernova event is established, with time delays between few days to several months (e.g. \cite{ouyedGRBs-1,ouyedGRBs-2} and references therein). This could be indicating the delayed onset of the quark-hadron phase transition, which can release sufficient energy to power the gamma-ray burst when neutron matter is burned to quark matter. Recently, this scenario has also been applied to detailed light-curve modelling of extended emission, afterglows and late-time X-ray flaring detected in several long GRBs by the Swift satellite \cite{staff2008}. Thermal nucleation timescales are also relevant to the proto-NS phase after the supernova explodes, when the appearance of a quark phase can delay the collapse to black holes until after deleptonization is complete.  Updated calculations of the nucleation rate and timescale should take into account  modified expressions for the shear viscosity of neutron matter and kaon condensed matter. The earliest studies of nucleation used the framework of quantum nucleation with the WKB approximation for tunneling, obtaining nucleation times in cold $\beta$-stable and hyperonic matter that had large uncertainties ($\tau_{\rm nuc}$ is exponentially sensitive to the overpressure).
We are presently working to reduce some of these uncertainties using updated
transport coefficients in dense matter.

\item  {\bf The Deflagration-Detonation-Transition (DDT)}:  The question remains open as to whether the combustion will switch from a deflagration to a detonation. Given the rapid flame speed that we find ($v\sim 0.01$c) and the size of the NS $R\sim 10$ km, the conversion timescale of order $10^{-3}$ seconds is of the same order as for the development of well-known flame instabilities - the Landau-Darrehius and Rayleigh-Taylor instabilities. Therefore, we have good reasons to expect that the laminar flow analysis in 1D is insufficient and turbulent flame burning must arise in 2D. Given this, it would seem likely to see a substantial increase in burning speed, but results from high-resolution multi-dimensional simulations are needed to make any conclusions. However, we note that in either case a second explosion would be realized; either the detonation explodes the outer-layers of the parent NS, or, the interface halts (due to the newly discovered deleptonization instability) and the outer-layers fall back onto the SQM core resulting in a core-bounce explosion.  The energy of the explosion is then due to either to chemical energy (the detonation), or to gravitational energy from core-collapse.

\item {\bf Temperature effects on SQM burning}:

Another important discovery was the effect of temperature on the combustion to quark matter. Not only can cooling change the properties of the interface, such as width and speed, but can quench the burning process entirely.
Thus, it is essential to include temperature increase when solving the jump conditions, as was shown in 
\cite{2010PhRvC..82f2801N} (see also \S~6 in \cite{Niebergal-thesis-2011}). It is important to keep in mind that while the Bag model (equation-of-state) has been extremely successful at explaining observations, both in experiment and in Nature, it is only considered to be a rough approximation. QCD allows for more sophisticated models, some of which may change the dynamics of SQM burning.  The \codename ~could readily be used to test these models but is left as an avenue for future work. 

\end{itemize}

\section{Conclusion}

We have performed 1D numerical simulations of the burning of neutron matter to strange quark matter with consistent treatments of reactions, diffusion, and hydrodynamics. Our two new results, viz., the generation of a rapid conversion front and its possible instability due to cooling effects by neutrino emission are exciting, but only suggestive at this point. Clearly, more sophisticated
numerical work is required to verify our  findings in a multi-dimensional setting, but there is widespread interest in modelling QCD phase transitions inside compact stars/core-collapse supernovae and the early universe. Such
groups could gain insight from our results and perform additional simulations.

The \codename ~ code is publicly available, written in a modular form allowing it to be easily adapted to other problems of flame burning and propagation in non-premixed fluids. The explosion from the transition of hadronic to \textit{u,d,s}-quark matter is not yet proved from first principles, but there are hints from the 1D study that such an explosion is possible. If confirmed, this would validate the Quark-Nova model\footnote{Our findings hint at a first order phase transition; a second order does not release any latent heat, so it probably does not happen explosively. Some other counter arguments to the idea of a first order phase transition stems from observations. These seem to be telling us precisely that we need a strong first order, and we have mentioned several observational signals of this, but theoretically its hard to decide between first and second because QCD parameters are not so well known at high density.} \cite{2002A&A...390L..39O,2005ApJ...618..485K},
consequences of which include: i) nucleosynthesis of rare elements \cite{Jaikumar:2006qx},
explanations for; ii) gamma-ray bursts \cite{2002A&A...387..725O,2005ApJ...632.1001O};  
iii) ultra-high energy cosmic rays \cite{2005ApJ...626..389O}; 
iv) soft-gamma repeaters, anomalous x-ray pulsars, x-ray dim NSs
\cite{2004A&A...420.1025O,2007A&A...473..357O,2007A&A...476L...5N,2007A&A...475...63O,2010PhRvD..81d3005N},
and ; v) super-luminous supernova \cite{2008MNRAS.387.1193L,2012MNRAS.423.1652O,2012arXiv1202.2400O}. The
  validation of the Quark-Nova would also have  implications to Cosmology 
\cite{2009ApJ...702.1575O,2013RAA....13..435O,2013MNRAS.428..236O,2013arXiv1304.3715O}.

\bigskip
R. O. is grateful to the organizers of the CSQCDIII for an inspiring meeting.
This research is supported by a grant from the National Research and Engineering Council of Canada (NSERC).


\end{document}